\documentclass[11pt,aps,prd,nofootinbib,showpacs,superscriptaddress,preprint]{revtex4}
\usepackage[utf8]{inputenc}
\usepackage{epsfig}
\usepackage{amsmath}
\usepackage{slashed} 
\usepackage{amsfonts} 
\usepackage{amssymb}
\usepackage{color}
\usepackage[colorlinks,citecolor=blue]{hyperref}
\usepackage{tabularx}
\usepackage{titlesec} 
\usepackage{natbib}
\usepackage{lipsum}
\usepackage{slashed}
\usepackage{tikz}
\usetikzlibrary{snakes}
\begin{document}
	
	\title{Gauged $L_e-L_{\mu}-L_{\tau}$ symmetry, fourth generation, neutrino mass and dark matter}
	
	\author{Satyabrata Mahapatra}
	\email{ph18resch11001@iith.ac.in}
	\affiliation{Department of Physics, Indian Institute of Technology Hyderabad, Kandi, Sangareddy 502285, Telangana, India}
	
	\author{Rabindra N. Mohapatra}
	\email{rmohapat@umd.edu}
	\affiliation{Maryland Center for Fundamental Physics and Department of Physics, University of Maryland, College Park, Maryland 20742, USA.}
	
	\author{Narendra Sahu}
	\email{nsahu@phy.iith.ac.in}
	\affiliation{Department of Physics, Indian Institute of Technology Hyderabad, Kandi, Sangareddy 502285, Telangana, India}
	
	\begin{abstract}
		We present two models where the familiar leptonic symmetry $L_e-L_\mu-L_\tau$ is a gauge symmetry. We 
		show how anomaly cancellation constrains the allowed theories, with one of them requiring a fourth 
		sequential chiral standard model fermion generation and a second one with three generations, requiring gauging of 
		$(L_e-L_\mu-L_\tau) - (B_1-B_2-B_3)$ with $B_a$ representing the baryon number of the $a$th generation quarks. 
		Unlike  global $L_e-L_\mu-L_\tau$ models which always leads to inverted mass hierarchy for 
		neutrinos, the gauged version can lead to normal hierarchy. We show how to construct realistic models 
		in both the cases and discuss the dark matter candidate in both. In our model, the breaking of 
		$U(1)_{L_e-L_\mu-L_\tau}$ is responsible for neutrino mass via type-I 
		mechanism whereas the 
		real part of $U(1)_{L_e-L_\mu-L_\tau}$ breaking scalar field (called $\phi$ here) plays the 
		role of freeze-in dark matter candidate. Since $\phi$ is unstable, for it to qualify as dark matter, its 
		lifetime must be larger than the age of the Universe, implying that the relic of $\phi$ is generated through 
		freeze-in mechanism and its mass must be less than an MeV. We also discuss the possibility of explaining both muon and electron $g-2$ while being consistent with the DM relic density and lifetime constraints.

	\end{abstract}
	\maketitle{}
	
	\section{INTRODUCTION}\label{intro}
	In spite of all the recent developments in the understandings of physics at the electro-weak scale, we 
	still have not found a complete solution to the puzzles of particle dark matter, neutrino masses, matter-antimatter 
	asymmetry etc. The solution to either of these problems indicates physics beyond the standard model (SM). The 
	observations of neutrino oscillation~\cite{Fukuda:1998mi,Ahmad:2001an,Abe:2011fz,An:2012eh,Ahn:2012nd} have 
	confirmed that neutrinos have nonzero but tiny masses. The oscillation data, though only sensitive to the 
	difference in mass-squareds and not the absolute mass scale of neutrinos; the stringent constraint for the 
	sum of neutrino masses is $\sum m_{\nu}\leq 0.12$ eV from  cosmological studies \cite{Tanabashi:2018oca,Vagnozzi:2017ovm,Giusarma:2016phn,Giusarma:2018jei}. 
	
	The astrophysical and cosmological observations of the galaxy rotation curve, gravitational lensing, large 
	scale structure formations and Cosmic microwave background radiation(CMBR) {\it etc.} provide ample evidences 
	for the existence of dark matter (DM)~\cite{Bertone:2004pz,Feng:2010gw}. The latest data from the 
	WMAP~\cite{Hinshaw:2012aka} and PLANCK~\cite{Aghanim:2018eyx} collaborations suggest that DM gives rise to 
	$26.8$\% of the total energy density of the present Universe.  But SM has no explanation for the origin of 
	the DM.
	
	Similarly the anomalous magnetic moment of the leptons (particularly the muon and electron) play an important role in the establishment of the SM and continue to act as a long-standing test of the SM while putting rigorous constraints on the beyond SM (BSM) theories. 
	Understanding the anomalous magnetic moment of leptons has received a lot of attention in the literature as it challenges the phenomenological success of the SM~\cite{Aoyama:2012wk,Aoyama:2019ryr,Czarnecki:2002nt,Gnendiger:2013pva,Davier:2017zfy,Keshavarzi:2018mgv,Colangelo:2018mtw,Hoferichter:2019mqg,Davier:2019can,Keshavarzi:2019abf,Kurz:2014wya,Melnikov:2003xd,Masjuan:2017tvw,Colangelo:2017fiz,Hoferichter:2018kwz,Gerardin:2019vio,Bijnens:2019ghy,Colangelo:2019uex,Blum:2019ugy,Colangelo:2014qya,Aoyama:2020ynm,Borah:2021khc,Crivellin:2018qmi}. The muon anomalous magnetic moment, $a_\mu$ = $(g - 2)_\mu/2$ has been measured recently by the E989
	experiment at Fermi lab showing a discrepancy with respect to the theoretical prediction of the Standard
	Model ($
	a^{\rm SM}_\mu = 116 591 810(43) \times 10^{-11}
	$). This measurement 
	$a^{\rm FNAL}_\mu = 116 592 040(54) \times 10^{-11}$ 
	when combined with the previous Brookhaven determination of
	$a^{\rm BNL}_\mu = 116 592 089(63) \times 10^{-11}$
	leads to a 4.2 $\sigma$ observed excess of
	$\Delta a_\mu = 251(59) \times 10^{-11}$ \cite{Abi:2021gix}. Here it is worth noting that, due to the non-perturbative character of the low energy strong interaction, the uncertainty in $a^{\rm SM}_\mu$ is mostly dominated by hadronic vacuum polarization (HVP) contributions. These contributions are calculated from data-driven approaches, utilizing measured $e^{+} e^{-} \to {\rm hadrons}$  data or from Lattice QCD. Results from various lattice groups are combined using a conservative procedure to give an average value to a leading order(LO) as $a^{\rm LO~HVP}_\mu=711.6(18.4)\times10^{-10}$\cite{Aoyama:2020ynm}. The most recent result from Lattice QCD with higher precision is from BMW-20 analysis~\cite{Borsanyi:2020mff} which gives $a^{\rm LO~HVP}_\mu=707.5(5.5)\times10^{-10}$. Similarly earlier measurements of HVP using $e^{+}e^{-} \to \pi^{+}\pi^{-}$ gives $a^{\rm HVP}_\mu= 6845(40)\times 10^{-11}$~\cite{Aoyama:2020ynm} and
	the same has been measured with a greater precision by the recent CMD-3 experiment to be $a_\mu$~\cite{CMD-3:2023alj}.  Though these observations help to reduce the tension between the experimental value of $a_\mu$ and its Standard Model prediction, they are far from conclusive, and the possibility of new physics driving this anomaly remains.
		Similarly anomalous magnetic moment for electron remains an open-question and the measurement using Rubidium atomic interferometry gives us $+$ve value of $\Delta a_e$ with $1.6\sigma$ discrepancy with SM predicted value \cite{Morel:2020dww}\footnote{ There is still an ambiguity in the sign of $\Delta a_e$ as from Cesium atomic interferometry, \cite{Parker:2018vye}, $\Delta a_e$ has $2.4 \sigma $ discrepancy with SM : $(\Delta a_e)_{\rm Cs} = (- 87 \pm 36) \times 10^{-14}$ which is $-$ve. Here we consider the recent measurement which depicts positive $(g-2)_e$ only. } 
	{\it i.e.}$(\Delta a_e)_{\rm Rb} = ( 48 \pm 30) \times 10^{-14}.
	\label{del_ae}$ 
	
	Also in the SM the number of fermion generations is one of the unresolved questions as there is no answer to 
	why only three generations of chiral fermions are observed. Therefore it is obvious to think whether there 
	exist additional families of quarks and leptons~\cite{Frampton:1999xi,Holdom:2006mr}. Indeed certain extensions 
	of SM suggest a particular family structure and some theories suggest an even number of fermion 
	generations~\cite{Hou:2008xd,Erler:1996zs,Chaudhuri:1994cd}. There are arguments favouring the existence of 
	the $4^{th}$ generation in the context of flavour democracies ~\cite{Sultansoy:2006dg,Datta:1992qd,Celikel:1994cu} 
	disfavouring the existence of $5^{th}$ generation.
	
	Considering the fact that symmetries have played a fundamental role in constructing the SM, it is natural to 
	ask whether they may help to improve our insight into the above issues. Among many local and global symmetries 
	discussed in the leptonic sector, $B-L$ where $B$ and $L$ represent the total baryon and lepton number, $L_\alpha-L_\beta$ where $\alpha,\beta = e,\mu,\tau$,  $L_e-L_\mu-L_\tau$, are some prominent ones. In this paper 
	we focus on the last one and propose a new class of models where it is local symmetry.
	
	We note that attention in the past has focussed  on $L_e-L_\mu-L_\tau$ only  as global leptonic symmetry. It has been extensively discussed in the literature  in connection with neutrino mass hierarchy (see for example ~\cite{Barbieri:1998mq, Joshipura:1998kg, Mohapatra:1999zr,Kitabayashi:2000nq,Lavoura:2000kg,Babu:2002ex,Goh:2002nk, Petcov:2004rk}) and large atmospheric mixing and even as a way to understand sterile neutrinos!~\cite{Mohapatra:2001ns,Lindner:2010wr}. A unique prediction of this is the inverted mass hierarchy of neutrinos and an atmospheric mixing angle of order one though  not necessarily maximal. The exact symmetry limit of such theories are however not acceptable since they predict zero  $\theta_{13}$ and no solar oscillation. One therefore has to include symmetry breaking terms. There is however no guidance as to how to choose the latter.  Only after making some assumptions about symmetry breaking terms,  one can predict mixing angles such as the $\theta_{13}$ as well as solar mixing $\theta_{12}$. The two questions that then arise are: (i) Is $L_e-L_\mu-L_\tau$ as a symmetry of the leptonic sector ruled out if the neutrino mass hierarchy is found from experiments to be normal and (ii) is there any way to make the symmetry breaking terms less arbitrary in analysing the neutrino masses? 
	
	In this note we show that if we gauge the $L_e-L_\mu-L_\tau$ symmetry, the resulting theories are much more restrictive (i) first due to cancellation of gauge anomalies and (ii) due to the fact that only certain forms of symmetry breaking terms are allowed, both due to renormalizability of the theory. One can show that there exist three viable ways to satisfy the anomaly conditions in a simple manner; thus leading to three different models. The model-I is a four generation version of the standard model with the gauge symmetry being $L_e-L_\mu-L_\tau+L_4$ with $L_4$ being the lepton number associated with the 4th generation of leptons.  
	In model-II, it is $(L_e-L_\mu-L_\tau )- (B_1-B_2-B_3) $ symmetry with only  three generations, where $B_a$ refers to the baryon number of the $a$th generation of quarks before SM symmetry breaking. And the gauge symmetry of model-III is $B+3(L_e-L_\mu-L_\tau)$ with three generations of fermions. 
 We note that the third possibility was pointed out in ~\cite{Heeck:2012cd} where the spontaneous breaking of the $B+3(L_e-L_\mu-L_\tau)$ symmetry in a type-I seesaw framework was shown to generate light neutrino mass matrices with approximate symmetries $L_e$ or $L_e-L_\mu-L_\tau$. While the remnant  symmetry $L_e$ predicts normal hierarchy(NH) of neutrino masses, the $L_e-L_\mu-L_\tau$ along with an additional $Z_2$ symmetry under which one of the RHNs is odd, predicts inverted hierarchy(IH) spectrum of neutrino masses. In the latter case, the $Z_2$ odd RHN is considered to be a viable DM candidate.  
	  
To the best of our knowledge, the first two models were not noticed in the literature. The model-I has a pure leptonic  gauge symmetry whereas the gauge symmetries of model-II and III arise from combinations of baryon and lepton numbers. The model-II and III have different symmetry structure though they share similar particle content\footnote{The two models have different number of scalars but they have same fermion content notably SM with three RHNs.}. As a result they lead to different phenomenology.  Here we analyse  model-I and II and obtain the following new results: (i) we find that there is enough structure in the models to reproduce observed quark and lepton mixings and discuss how  in certain parameter domains, they can lead to a dark matter candidate.  We also see that if one promotes the $L_e-L_\mu-L_\tau$ symmetry to be a local symmetry, then both normal and inverted hierarchies of neutrino masses are possible. 
	(ii) Secondly,  we find that in either model the real part of the complex scalar field that breaks the gauge symmetry can be a viable DM candidate whose relic density is generated by the freeze-in mechanism. A similar result was noted in ~\cite{Mohapatra:2020bze} for a $B-L$ model case where the real part of the scalar field that breaks $B-L$ though unstable, has a long enough lifetime to satisfy astrophysical constraints on unstable dark matter. We find that the dark matter has to have a mass less than an MeV to be consistent with all the relevant constraints. (iii) Finally, we show the viable parameter space in the plane of new gauge coupling and gauge boson mass that can explain the anomalous magnetic moment of muon as well as electron while being consistent with the life time and relic density constraints on the freeze-in dark matter.
	
	The rest of the paper is organised as follows. In section \ref{anomaly}, we discuss the chiral gauge anomaly in $U(1)_{L_e-L_{\mu}-L_{\tau}}$ extension of SM and propose two possible extensions of SM from the anomaly cancellation requirement. In section~\ref{model-A}, we describe the first model with gauged $U(1)_{L_{e}-L_{\mu}-L_{\tau}}$ extension of the SM along with all the relevant constraints on the $4^{\rm th}$ generation quarks and leptons which are introduced for anomaly cancellation. Similarly in section~\ref{model-B}, we discuss the details of the second model which has three generations and $U(1)_{(L_e-L_\mu-L_\tau )- (B_1-B_2-B_3)}$ gauge symmetry. Then, in section~\ref{nu_mass}, we discuss the generation of light neutrino mass for both the models. In section~\ref{darkmatter}, we discuss the solution to the DM problem of the Universe in both the proposed models along with all the relevant constraints and phenomenology. In section~\ref{g-2anomaly}, we discuss the possibility of explaining the anomalous magnetic moments of electron and muon via the contribution from the new gauge boson loop. We discuss the possible prospects of detection of the models briefly in section~\ref{collider_sign} and finally conclude in section~\ref{conclusion}.
	
	\section{Anomalous gauged $U(1)_{L_e-L_\mu-L_\tau}$ symmetry and two possible extensions of the SM}\label{anomaly}
	Since the SM is a chiral gauge theory, the anomalies arising from the triangle diagrams involving the gauge currents is proportional 
	to:
	\begin{equation}
		\mathcal{A}=~Tr\Big(T_a[T_b,T_c]_+\Big)_R-~Tr\Big(T_a[T_b,T_c]_+\Big)_L\,,
	\end{equation}
	where $\mathcal{A}$ stands for the anomaly coefficient. The $T$'s denote the generators of the gauge groups and $R$ and $L$ 
	represents the interactions of right and left chiral fermions with the gauge bosons. By construction, each 
	generation of the SM is anomaly free. 
	
	Let us augment the SM by including a gauged $U(1)_{{L_e-L_\mu-L_\tau}}$ symmetry, where $L_e$,$L_\mu$ and $L_\tau$ 
	are the respective family lepton numbers. For convenience, from now on we will denote $L_e-L_\mu-L_\tau=X$. In such an extension 
	of the SM, without adding any new particle, the non-trivial triangle anomalies arise from the combinations: 
	$SU(2)^2_{L} \times U(1)_{X}$, $U(1)^2_{Y} \times U(1)_{X}$ and $U(1)^3_{X}$. The corresponding anomaly coefficients can 
	be calculated to be 1,-1/2 and 1 respectively. In this case, we find that the triangle anomalies can be cancelled by introducing an extra generation of chiral leptons $L_4=(N, E)_L^T,~E_{R}$ to the SM with $U(1)_X$ charge $+1$. But as we know that each individual fermion generation 
	of the SM is free of $SU(3)_C \times SU(2)_L \times U(1)_Y$ gauge anomalies. Therefore, we further introduce a new generation of 
	quarks $Q_4=(U, D)_L^T,~U_R,D_R$ to the SM for this purpose. Here $Q_4$ and  $L_4 $ are the fourth generation left handed quarks and leptons which transforms as doublets under $SU(2)_{L}$, while $U_R,D_R$ are the corresponding right handed quarks and $E_R$ is the corresponding charged lepton which are $SU(2)_{L}$ singlets. The anomaly free new $U(1)$ is actually $U(1)_{{L_e-L_\mu-L_\tau+L_4}}$ Thus we end up with a model (to be called model-A) having four generations of quarks and leptons with gauged $U(1)_{{L_e-L_\mu-L_\tau}}$ symmetry in the observed lepton sector. To this model, we will also add four SM singlet right handed fermions  denoted by $N_a$ for implementing type I seesaw without affecting the anomaly discussion. The detailed phenomenology of model-A will be discussed in section \ref{model-A}.
	
	In the same vein, we also attempt to gauge $X'=(L_e-L_\mu-L_\tau) - (B_1-B_2-B_3)$, where $B_a$ is the baryon number of the $a$th generation quarks, using the SM particles and three generations. We find that the only non-trivial triangle anomaly arises from $U(1)_{X'}^3$ and can be trivially shown to be 1. We cancel this anomaly by introducing three generations of right handed neutrinos (RHNs) $N_e, N_\mu, N_\tau$ 
	with $X'$ charges 1,-1,-1 respectively. Thus we end up with an anomaly free model (to be called model-B) with SM + 3 RHNs. The detailed phenomenology of model-B will be discussed in section \ref{model-B}.
	
	We note parenthetically that similar conclusions about anomaly cancellation of $U(1)_{L_e-L_\mu-L_\tau}$ hold for extended electroweak gauge groups such as the left-right group $SU(2)_L\times SU(2)_R\times U(1)_{B-L}$. Here we focus only on the extension of the SM gauge group.

	\section{The model-A: Four generations of quarks and leptons, neutrino mass and dark matter }\label{model-A}
	As discussed in section \ref{anomaly}, the model-A constitutes four generations of quarks and leptons. The gauge symmetry of the 
	model-A is $SU(3)_C \times SU(2)_L \times U(1)_Y \times U(1)_X$, where $X=L_e-L_\mu-L_\tau+L_4$. To generate light neutrino masses 
	through type-I seesaw we further introduce four generations of heavy right-handed neutrinos $N_{\alpha}$, $\alpha=e,\mu,\tau$ and $N_{4}$ with having $U(1)_X$ charges 1,-1,-1,1 respectively. Here it is worth mentioning that for anomaly cancellation the $U(1)_X$ charge assignment of the fourth generation leptons becomes identical to that of the first generation {\it i.e.}, the electron family. So to segregate the fourth generation lepton from other three generation of SM-leptons, we impose an additional discrete $Z_2$ symmetry under which $L_4 =(N , E)^T_L,~E_R$ and the gauge singlet RHN $N_{4}$ are odd while other three generations of leptons are even. Similarly the fourth family of quarks are chosen to be odd under the $Z_2$ symmetry so that unwanted mixing between fourth family with other three generations of quarks can be avoided. To give masses to up type and down type quarks we introduce two Higgs doublets 
	$H_1$ and $H_2$. A second $Z'_2$ symmetry is also introduced under which all down type quarks, all right handed charged leptons including 4th generation and $H_2$ are odd, while all other particles are even.   
	
	To break the $U(1)_X$ symmetry we introduce a SM singlet scalar $\Phi$ with $U(1)_X$ charge -2. The particle content and their charge assignments are shown in Table.\ref{table}. 
	
	\begin{table}
		\begin{center}
			\begin{tabular}{|c||c|c|c||c|c|c||c|c| c|c||c|c|c||c|c|c|}
				\hline
				\bf Fields & $Q_i$  & $u_{Ri}$ & $d_{Ri}$ & $Q_4$ & $U_R$ & $D_R$ & $L_e,e_R$ & $L_\mu, \mu_R$ & $L_\tau, \tau_R$ & $N_{e}, N_\mu, N_\tau$ & $L_4$ & $E_R$ & $N_4$ & $H_1$ & $H_2$ & $\Phi$ \\ 
				\hline
				\hline
				$SU(3)_c$ & 3 & 3 & 3 & 3 & 3 &3& 1,1 & 1,1 & 1,1 & 1,1,1 & 1 & 1& 1& 1& 1 & 1\\
				\hline
				$SU(2)_L$ & 2 & 1 &1 & 2 & 1 & 1& 2,1 & 2,1 & 2,1 &1,1,1 & 2 & 1&1&2&2&1\\
				\hline
				$U(1)_Y$ & 1/6 & 2/3 & -1/3 & 1/6 & 2/3 & -1/3 & -1/2,-1 & -1/2,-1 & -1/2,-1 & 0,0,0 & -1/2 & -1 & 0&1/2&1/2&0 \\
				\hline
				\hline
				$U(1)_{X} $ & 0 & 0 & 0 & 0 & 0& 0 & 1,1 & -1,-1 & -1,-1& 1,-1,-1 & 1 & 1 & 1&0&0&-2\\
				\hline
				$Z_2$ & + & + & + & - & - &- & +,+ & +,+ & +,+ & +,+,+ & - & - &- & + & + & +\\
				\hline
				$Z'_2$ & + & + & - & + &  + &- & +,- & +,-& +,- & +,+,+ & +& - & + & + &- & +\\
				\hline
			\end{tabular}   
			\caption{Particle spectrum of the model with their quantum numbers under the imposed symmetry. Here $i=1,2,3$ and $X=L_e-L_\mu-L_\tau$.}
			\label{table}
		\end{center}
	\end{table}

	The scalar potential involving $\Phi$, $H_1$ and $H_2$ is given by:
	\begin{align}
		\mathrm{V}(H_1,H_2,\Phi) &= -M^2_1 H^\dagger_1 H_1 + \lambda_1 {(H^\dagger_1 H_1)^2} - M^2_2 H^\dagger_2 H_2 + \lambda_2 {(H^\dagger_2 H_2)^2}-\mu^2(H^\dagger_1 H_2 + H^\dagger_2 H_1)\nonumber\\&+ \lambda_{12}(H^\dagger_1 H_1)(H^\dagger_2 H_2)  +\lambda'_{12}(H^\dagger_1 H_2)(H^\dagger_2 H_1) + \frac{\lambda''_{12}}{2}\left[(H^\dagger_1 H_2)^2+(H^\dagger_2 H_1)^2\right]\nonumber\\&- M^2_{\Phi} \Phi^\dagger \Phi + 
		\lambda_{\Phi} {(\Phi^\dagger \Phi)^2}+\lambda_{1 \Phi} (H^\dagger_1 H_1)(\Phi^\dagger \Phi)+\lambda_{2 \Phi} (H^\dagger_2 H_2)(\Phi^\dagger \Phi) \,.
		\label{vscalar}
	\end{align}
	We assume that all the parameters in potential $V(H_1,H_2,\Phi)$ to be real. We also make an assumption that $\lambda_{1 \Phi}=\lambda_{2 \Phi}=0$. We will comeback to this point in section \ref{darkmatter} while discussing about the viability of real part $\Phi$ being a dark matter candidate.

	After the Electro-weak symmetry breaking, the physical mass spectrum contains two CP-even Higgs bosons $h_{1,2}$, one CP-odd Higgs $\eta$, and the charged Higgs pair $H^{\pm}$. Hence, the original Higgs doublets can be written in terms of the above mentioned fields along with the Goldstone bosons as:
	\begin{equation*}
		H_{1} =\frac{1}{\sqrt{2}}\begin{pmatrix}
			\sqrt{2}(c_\beta G^+-s_\beta H^+)\\ {c_\beta v-s_\alpha h_{1}+c_\alpha h_2 + i(c_\beta G^0-s_\beta \eta)}
		\end{pmatrix}~~~{\rm ,}~~~~
		H_{2} =\frac{1}{\sqrt{2}}\begin{pmatrix}
			\sqrt{2}(s_\beta G^++c_\beta H^+)\\ {s_\beta v+c_\alpha h_{1}+s_\alpha h_2 + i(s_\beta G^0+c_\beta \eta)}
		\end{pmatrix}
	\end{equation*}
	and $$
	~~ \Phi =\frac{v_\phi+\phi+i \xi}{\sqrt{2}}.$$
	
	\noindent Here $s_\beta \equiv \sin \beta$, $c_\beta\equiv\cos\beta$ and $\alpha$ is the mixing between $h_1$ and $h_2$ and the ratio of the vev of the two Higgs fields is defined by $\tan\beta$. We recognise $h_1$ as the SM like Higgs with mass $M_{h_1}$ and $h_2$ as the heavy CP-even Higgs with mass $M_{h_2}$.

	It is worth mentioning that $\mu^2 H_1^\dagger H_2$ explicitly breaks $Z'_2$ in order to avoid the domain wall problem 
	at the electroweak scale. 
	
	The mass terms of the charged leptons consistent with the imposed symmetry are:
	\begin{align}
		\mathcal{L} \supset & -y_{ee}\overline{L_{e}} H_2 e_R -y_{\mu \mu}\overline{L_{\mu}} H_2 \mu_R-y_{\tau \tau}\overline{L_{\tau}} H_2 \tau_R\nonumber-y_{\mu \tau}\overline{L_{\mu}} H_2 \tau_R-y_{ \tau\mu}\overline{L_{\tau}} H_2 \mu_R-y_{EE}\overline{L_{4}} H_2 E_R+ ~h.c.
	\end{align}
	and thus the SM charged lepton mass matrix $\mathcal{M}_l$ has the texture :
	\begin{equation}
		\mathcal{M}_\ell= \begin{pmatrix}
			y_{ee}\langle H_2 \rangle && 0 && 0 \\
			0 && y_{\mu \mu}\langle H_2 \rangle && y_{\mu \tau}\langle H_2 \rangle  \\
			0 && y_{\tau \mu}\langle H_2 \rangle  && y_{\tau \tau}\langle H_2 \rangle\\ 
		\end{pmatrix}
		\label{clm}
	\end{equation}
	
	As can be seen from the charged lepton mass matrix texture in \ref{clm}, it has off-diagonal terms which implies lepton flavour violation. CMS collaboration has already reported an event excess in a flavor-violating Higgs decay $h \rightarrow \mu \tau$, and 
	it suggests that the best fit value for the branching fraction is
	$\mathcal{B}r (h \rightarrow \mu \tau ) = (0.84^{+0.39}_{-0.37})$\%. But the constraint on the branching fraction is
	$\mathcal{B}r(h \rightarrow \mu \tau )<1.51$\% which has been set at 95\% C.L. . This essentially limits 
	the $\mu$-$\tau$  Yukawa couplings to be less than $3.6\times 10^{-3}$\cite{CMS:2015qee}. In this case the final state was a sum of $\mu^+ \tau^-$and $\mu^- \tau^+$ and the deviation from the SM prediction is $2.4\sigma$. In addition to it the results of the ATLAS experiment has also shown $\mathcal{B}r(h \rightarrow \mu\tau ) = (0.77 \pm 0.62) \%$ which is consistent with the CMS result within $1 \sigma$ ~\cite{ATLAS:2015cji}. 
	
	Here it is worth mentioning that, because of the imposed $Z_2$ symmetry under which the 4th generation fermions are odd while all other particles are even, they get decoupled from the 3 SM generations.

	The Lagrangian for quark masses in our setup is given by:
	
	\begin{eqnarray}
		\mathcal{L} \supseteq -Y^d_{ij} \overline{Q_i} H_2 d_{Rj}-Y^u_{ij}\overline{Q_i} \tilde{H_1} u_{Rj}-Y^d_{4}\overline{Q_4} H_2 D_{R}-Y^u_{4}\overline{Q_4}\tilde{H_1} U_{R}
	\end{eqnarray}

	The Lagrangian involving fourth generations of leptons consistent with the extended symmetry can be written as:
	\begin{align}
		\mathcal{L} ~\supset~~ & \overline{L_4}\, i \gamma^\mu D_{\mu} L_4 + \overline{E_R} i \gamma^\mu D'_{\mu} E_R + 
		{\big| D_{\mu} \Phi \big|}^2 + \overline{N_4} i \gamma^\mu D''_\mu N_4  - f_4 \Phi \overline{(N_4)^c} N_4 -{Y_{4}\overline{L_4} {H_1} N_{4} }+h.c
		\label{Lagr}
	\end{align}
	where 
	\begin{equation*}
		D_{\mu}=\left( \partial_\mu + ig T^i W^i_{\mu} + ig'YB_{\mu} + ig_{_X} X (Z_{_X})_\mu  \right)
	\end{equation*}
	\begin{equation*}
		D'_{\mu}=\left( \partial_\mu + ig'YB_{\mu} + ig_{_X} X (Z_{_X})_\mu  \right) \nonumber\\.
	\end{equation*} 
	and 
	\begin{equation*}
		D''_{\mu}=\left( \partial_\mu + ig_{_X} X (Z_{_X})_\mu  \right) \nonumber\\.
	\end{equation*} 
	The $g_{_X}$ is the gauge coupling associated with $U(1)_X$ and $Z_{_X}$ is the corresponding gauge boson.

	\subsection{Constraints on fourth generation lepton and quark masses}
	In the context of a fourth generation, the most obvious restrictions that should be taken care of are the invisible width of $Z$ boson, generational mixings, precision electroweak measurements and the direct search bounds~\cite{He:2001tp,Kribs:2007nz,Eberhardt:2010bm,Erler:2010sk,Djouadi:2012ae,Bar-Shalom:2011lgb}. There exist bounds on the sequential fourth generation particles from collider searches \cite{Tanabashi:2018oca}. If the $4^{th}$ generation Dirac neutrino decays to other particles then it has to be heavier than
	90.3 GeV and if it is a Majorana particle then it's mass should be greater than $80.5$GeV. However if it is stable and long lived at LEP, then from the invisible $Z$-boson decay width the only bound is $m_N > 45.0$ GeV if Dirac and $m_N > 39.5$ GeV if Majorana. However if the fourth generation neutrino have both Dirac and Majorana mass giving rise to two Majorana neutrino mass eigen states, then taking into account the mixing between them, the lower mass bound can be further reduced to 33.5 GeV~\cite{Carpenter:2010sm}.  For the charged leptons, the bound is $m_E > 102.6$ GeV if it is stable and $m_E > 100.8$ GeV if it decays to other particles like $\nu W$.
	
	Similarly, there exist lower bounds for the sequential $4^{th}$ generation quark masses from the direct searches at the TeVatron. The limits on mass of the fourth generation down type quark is: $m_D > 190$ GeV for quasi-stable D, $m_D > 755$ GeV from neutral-current decays and $m_D > 880$ GeV from charged-current decays. And the limits on the mass of the fourth generation up type quark is: $m_U > 1160$ GeV from neutral-current decays and $m_U > 770$ GeV from charged-current decays. 
	
	It should be noted here that, in the experimental searches it is usually assumed that the mode under study has 100\% branching fraction and these heavy quarks are unstable. But these exclusion limits can be mitigated if the branching fraction or lifetime depends on the mixing angles and the mass splittings~\cite{Hung:2007ak}. So if one heavy quark could even become stable by virtue of some discrete symmetry such that its mixing with the lighter SM quarks is extremely small, then these constraints can be significantly loosened.
	
	In our set up, the fourth generation quarks and leptons are odd under a $Z_2$ symmetry and hence they are segregated from the three light generations. As a result the 4th generation fermions escape the constraints coming from precision electroweak measurements and collider searches  and can produce distinctive signals at collider experiments different from those for sequential 4G models~\cite{Murayama:2010xb,Lee:2012xn}.

	Both LHC collaborations (CMS and ATLAS) independently excluded Higgs mass in the range $120$ GeV $< M_h < 600$ GeV in case of 
	a sequential fourth generation fermions with one Higgs doublet. Therefore, the observed SM Higgs mass: $M_h=124.5$ GeV hints to 
	a beyond single Higgs doublet model as is the case we have in the proposed model. Two Higgs doublet models with fourth family are consistent with electro-weak precision constraints~\cite{Bar-Shalom:2011lgb, Hashimoto:2010at} as well as with the bounds from the Higgs signal strengths~\cite{Das:2017mnu}. 
		Here we briefly discuss the constraints from electroweak precision data on the fourth generation fermions and the additional Higgs doublet in our model.  
	
	In general, the contribution to the oblique parameters ($S,T,U$) from the fourth generation fermions ($\Delta S_f,\Delta T_f, \Delta U_f$) and the scalar doublet ($\Delta S_s, \Delta T_s, \Delta U_s$) are calculated with respect to the SM values and are bounded by\cite{Bar-Shalom:2011lgb,Gcollab}:
	
	\begin{eqnarray}
		\Delta S=S-S_{\rm SM}&=& 0.02 \pm 0.10\nonumber\\
		\Delta T=T-T_{\rm SM}&=& 0.07 \pm 0.12\nonumber\\
		\Delta U=U-U_{SM}&=&0.00 \pm 0.09
	\end{eqnarray}
	
	It is worth mentioning here that in the Higgs sector the parameters in the range $|\lambda''_{12}|<\pi$ and $0.5<\tan\beta<5$ with the mass range  $114~{\rm GeV}<M_{h_1}<1$~TeV,  and $M_{h_1}<M_{h_2}<1.5$ TeV, $300$ GeV$<M_{H^{\pm}}<1$ TeV, along with the mass ranges of the fermions $256$ GeV $<M_{U}<552$ GeV, $255$ GeV $<M_{D}<552$ GeV, $100.8$ GeV $<M_{E}<1.23$ TeV and $ 90.3$ GeV $<M_{N}<1.23$ TeV are consistent with the Electro-weak observable constraints as mentioned in the above equation \cite{Hashimoto:2010at}\footnote{These conclusions remains valid even with $M_{h_1}=125$ GeV\cite{Bar-Shalom:2012vvt}.}.  
	\subsection{Constraints on 4th generation neutrino in our set up}
	After $U(1)_X$ symmetry breaking by the vacuum expectation value (vev) of $\Phi$, $N_4$ acquires a Majorana mass 
	$f_4 \langle\Phi\rangle$. Similarly the electroweak symmetry breaking induces a Dirac mass 
	$Y_4 \langle H_1 \rangle$ to $N_4$. If $f_4 \langle\Phi\rangle >> Y_4 \langle H_1 \rangle$ then type-I
	see-saw mechanism comes into play and in the physical spectrum we get a relatively light 4th generation 
	neutrino with mass $M_{N} \simeq (Y^2_4 v^2_1)/(f_4 v_\phi)$. However, the fourth-generation neutrino must be massive 
	such that it does not contribute to invisible decay of $Z$ boson ( which is determined precisely from LEP measurements), 
	{\it i.e.} $M_N \geq \frac{M_Z}{2} $. In order to escape the strong limits on the number of active neutrinos and to satisfy 
	the present collider bounds, a proper tuning of the relevant parameters is required. In order to have $M_{N} \geq \frac{M_Z}{2}$ 
	we should have $f_4 v_\phi \leq 2 (Y^2_4 v^2_1)/M_Z$ which implies that mass of $N_4$ should be several hundreds of GeVs and 
	not much larger than the electroweak scale.
	
	From Eq,\ref{Lagr}, the mass matrix for 4th generation neutrino can be written as :
	\begin{equation}
		-\mathcal{L}_{N}=	\frac{1}{2}
		\begin{pmatrix}
			\overline{(N_L)^c} &
			\overline{(N_4)}\\
		\end{pmatrix}
		{\boldsymbol{\mathcal{M}}}
		\begin{pmatrix}
			N_L \\
			(N_{4})^c \\ 
		\end{pmatrix}
		+h.c	
	\end{equation}
	where 
	\begin{equation}
		{\boldsymbol{\mathcal{M}}}=\begin{pmatrix}
			0 && m_D \\
			m_D && m_R \\ 
		\end{pmatrix}
	\end{equation}
	with $m_{D}=(Y_4 v_1)/\sqrt{2}$ and $m_R=(f_4 v_\phi)/\sqrt{2}$. This mass matrix can be diagonalised by an unitary transformation using the transformation matrix 
	\begin{equation*}
		U=\begin{pmatrix}
			i\cos(\theta) && \sin(\theta) \\
			-i\sin(\theta) && \cos(\theta) \\ 
		\end{pmatrix}
	\end{equation*} and we obtain the mass eigen values given by:
	\begin{equation*}
		m_1=\frac{1}{2}\big[\sqrt{4m^2_D+(m_R)^2}-m_R~\big]
	\end{equation*}
	\begin{equation}
		m_2=\frac{1}{2}\big[\sqrt{4m^2_D+(m_R)^2}+m_R~\big]	
		\label{mass_v}
	\end{equation} 
	with corresponding mass eigen states:
	\begin{equation*}
		N_{1}=-i\cos\theta (N_L) + i\sin\theta (N_4)^c
	\end{equation*}
	\begin{equation}
		N_{2}=\sin\theta (N_L) + \cos\theta (N_4)^c
		\label{masseigen}
	\end{equation}
	where the mixing parameter is : 
	\begin{equation}
		\theta= \frac{1}{2} \tan^{-1}\Big(\frac{2m_D}{m_R}\Big)
	\end{equation}

	
	\begin{figure}
		\includegraphics[scale=0.5]{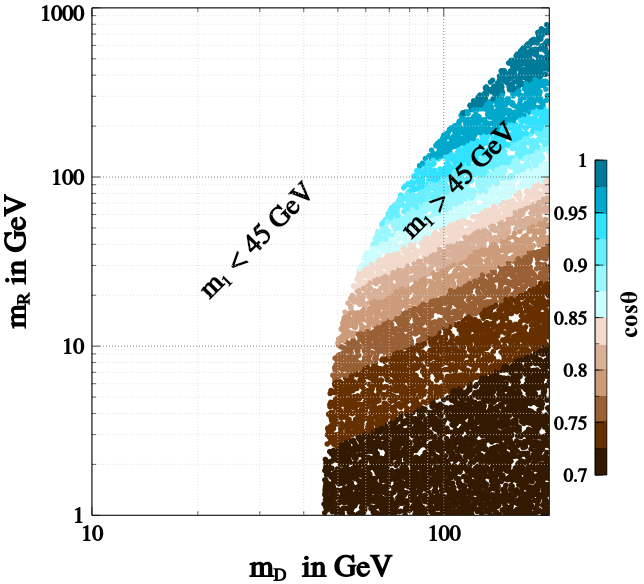}
		\caption{The parameter space in the plane of $m_D$ and $m_R$ that can give $m_1 > 45$ GeV. The colour code shows the values of the corresponding $\cos\theta$ values that governs the interaction strength of $N_1$ with $Z$ boson.}
		\label{n4mass}
	\end{figure}
	
	%
	
	In Fig.~\ref{n4mass}, we have shown the parameter space in the plane of $m_D$ and $m_R$ that can give $m_1 > 45$ GeV. The colour code shows the values of the corresponding $\cos\theta$ values that governs the interaction strength of $N_1$ with $Z$ boson as the weak neutral current for the fourth generation neutrinos can be written as :
	\begin{align}
		\mathcal{L}_{NC}\supseteq -\frac{g}{2\cos\theta_w}Z_{\mu}\big[&-\cos^2(\theta)\overline{N_1}\gamma^\mu \gamma^5 N_1-2i\cos\theta\sin\theta\overline{N_1}\gamma^\mu N_2-\sin^2(\theta)\overline{N_2}\gamma^\mu \gamma^5 N_2\big]
	\end{align}
	Thus, the coupling of the neutrinos with the $Z$ boson varies with the mixing angle and the coupling of the fourth generation 
	neutrino $N_1$ to $Z$ boson increases as the mixing angle decreases. From Fig.~\ref{n4mass}, it is clear that the parameter space consistent with $N_1>45$ GeV also implies its coupling with the $Z$ boson is large.

	\section{The model-B: Gauged $(L_e-L_\mu-L_\tau)-(B_1-B_2-B_3)$ symmetry, neutrino mass and dark matter }\label{model-B}
	As discussed in section \ref{anomaly}, the model-B represents a SM extension with three generations of right handed neutrinos (RHNs)
	$N_{\alpha}$, $\alpha=e,\mu,\tau$. The gauge symmetry of the model-B is $SU(3)_C \times SU(2)_L \times U(1)_Y \times U(1)_{X'}$, where 
	$X'=(L_e-L_\mu-L_\tau)-(B_1-B_2-B_3)$. We also extend the scalar sector of the SM model with two complex scalars $\Phi_1$ 
	and $\Phi_2$ with $U(1)_{X'}$ charges 2 and 2/3 respectively such that vacuum expectation values (vevs) of $\Phi_1$ and $\Phi_2$ 
	break the $U(1)_{X'}$ symmetry. We note 
	that the vev of $\Phi_1$ gives sub-eV masses to light neutrinos via type-I seesaw, while the vev of $\Phi_2$ is responsible 
	for quark masses. The particle content and their charge assignments under SM $\times U(1)_{X'}$ symmetry is given in Table.\ref{table}.
	
	\begin{table}
		\begin{center}
			\begin{tabular}{|c c||c|c|c|c|c|}
				\hline
				& Fields & $SU(3)_c$ & $SU(2)_L$ & $U(1)_Y$ & $U(1)_X'$\\ 
				\hline
				\hline
				& $L_e$, $e_R$ & 1,1 & 2,1 & -1/2,-1 & 1 \\
				\hline
				& $L_\mu, \mu_R$ & 1,1 & 2,1 & -1/2,-1 & -1\\
				\hline
				& $L_\tau,\tau_R$ & 1,1 & 2,1 & -1/2,-1 & -1\\
				\hline
				& $N_e$ & 1 & 1 & 0 & +1 \\
				\hline
				& $N_\mu$ & 1 & 1 & 0 & -1\\
				\hline
				& $N_\tau$ & 1 & 1 & 0 & -1\\
				\hline
				& $Q_1$, $U_{1R}$, $D_{1R}$ & 3,3,3 & 2,1,1 & $\frac{1}{6}$,$\frac{2}{3}$,-$\frac{1}{3}$,  & -$\frac{1}{3}$\\
				\hline
				& $Q_2$, $U_{2R}$, $D_{2R}$ & 3,3,3 & 2,1,1 & $\frac{1}{6}$,$\frac{2}{3}$,-$\frac{1}{3}$, & $\frac{1}{3}$\\
				\hline
				& $Q_3$,$U_{3R}$, $D_{3R}$  & 3,3,3 & 2,1,1 & $\frac{1}{6}$,$\frac{2}{3}$,-$\frac{1}{3}$, & $\frac{1}{3}$\\
				\hline
				& $\Phi_1$ & 1 & 1 & 0 & 2\\
				\hline
				& $\Phi_2$ & 1 & 1 & 0 & $\frac{2}{3}$\\
				\hline
				& $H$ & 1 & 2 & 1/2 & 0\\
				\hline
			\end{tabular}   
			\caption{Particles and their quantum numbers under the imposed symmetry: $SU(3)_C \times SU(2)_L\times U(1)_Y \times U(1)_{X'}$.}
			\label{table}
		\end{center}
	\end{table}
	
	The scalar potential involving $H$, $\Phi_1$ and $\Phi_2$ can be given as
	
	\begin{align}
		\mathrm{V}(H,\Phi_1,\Phi_2) &= -\mu^2 H^\dagger H + \lambda_H {(H^\dagger H)^2} - M_1^2 |\Phi_1|^2+ 
		\lambda_1 |\Phi_1|^4 - M_2^2 |\Phi_2|^2 + \lambda_{2} |\Phi_2|^4 \nonumber\\
		& + \lambda_{1H} (H^\dagger H)|\Phi_1|^2 +\lambda_{2H} (H^\dagger H)|\Phi_2|^2 + \lambda_{12} |\Phi_1|^2 |\Phi_2|^2\,. 
		\label{vscalar-B}
	\end{align}
	We assume $\lambda_{1H}=\lambda_{2H}=0$. The implication of this choice will be discussed in section \ref{darkmatter}. 
	
	The $U(1)_{X'}$ symmetry is broken by the vevs of $\Phi_1$ and $\Phi_2$, while the electroweak symmetry is 
	broken by the vev of H. The quantum fluctuations of fields around the minimum can be given as
	\begin{equation*}
		H=\begin{pmatrix}
			0\\ {(v + h)}/{\sqrt{2}}
		\end{pmatrix}~  {\rm ,}~~ \Phi_{1,2} =\frac{v_{1,2}+\phi_{1,2}+i \xi_{1,2}}{\sqrt{2}}\,.
	\end{equation*}

	The mass terms of the charged leptons consistent with the imposed symmetry are:
	\begin{align}
		\mathcal{L} \supset & -y_{ee}\overline{L_{e}} H e_R -y_{\mu \mu}\overline{L_{\mu}} H \mu_R-y_{\tau \tau}\overline{L_{\tau}} H \tau_R\nonumber\\
		&-y_{\mu \tau}\overline{L_{\mu}} H \tau_R-y_{ \tau\mu}\overline{L_{\tau}} H \mu_R - ~h.c.
	\end{align}
	and thus the SM charged lepton mass matrix $\mathcal{M}_l$ has the texture :
	\begin{equation}
		\mathcal{M}_\ell= \begin{pmatrix}
			y_{ee}\langle H \rangle && 0 && 0 \\
			0 && y_{\mu \mu}\langle H \rangle && y_{\mu \tau}\langle H \rangle  \\
			0 && y_{\tau \mu}\langle H \rangle  && y_{\tau \tau}\langle H \rangle\\ 
		\end{pmatrix}
		\label{clm-B}
	\end{equation}
	
	The Lagrangian describing quark masses in this set up is given by
	\begin{align}
		\mathcal{L} \supseteq & -Y^d_{ii} \overline{Q_i} H d_{iR}- Y^d_{12} \overline{Q_1} H d_{2R} \frac{\Phi_2^*}{\Lambda} -Y^d_{13} \overline{Q_1} H d_{3R} \frac{\Phi_2^*}{\Lambda} \nonumber\\
		& - Y^d_{21} \overline{Q_2} H d_{1R} \frac{\Phi_2}{\Lambda}- Y^d_{23} \overline{Q_2} H d_{3R} - Y^d_{31} \overline{Q_3} H d_{1R} \frac{\Phi_2}{\Lambda} - Y^d_{32} \overline{Q_3} H d_{2R}\nonumber\\
		& -Y^u_{ii}\overline{Q_i} \tilde{H} u_{Ri} - Y^u_{12}\overline{Q_1} \tilde{H} u_{2R} \frac{\Phi_2^*}{\Lambda} - Y^u_{13}\overline{Q_1} \tilde{H} u_{3R} \frac{\Phi_2^*}{\Lambda} \nonumber\\
		& - Y^u_{21}\overline{Q_2} \tilde{H} u_{1R} \frac{\Phi_2}{\Lambda} - Y^u_{23}\overline{Q_2} \tilde{H} u_{3R} - Y^u_{31}\overline{Q_3} \tilde{H} u_{1R} \frac{\Phi_2}{\Lambda} - Y^u_{32} \overline{Q_3} \tilde{H} u_{2R}\,.
		\label{quarkmassB}
	\end{align}
	It is clear that this Lagrangian will give a realistic mass matrix for quarks.

	\section{Light neutrino mass}\label{nu_mass}
	As stated in model-A and model-B, the sub-eV masses of light neutrinos can be generated via the canonical seesaw mechanism~\cite{Minkowski:1977sc,GellMann:1980vs,Mohapatra:1979ia,Yanagida:1980xy}. 
	Note that the charge assignment of $(L_e, e_R)$, $(L_\mu, \mu_R)$ and $(L_\tau, \tau_R)$ under $U(1)_X$ symmetry in model-A and 
	under $U(1)_{X'}$ symmetry in model-B is exactly same and is given as $1,-1,-1$ respectively. Similarly, the RHNs $N_e$, $N_\mu$ and $N_\tau$ carry same charges under $U(1)_X$ symmetry in model-A and under $U(1)_{X'}$ symmetry in model-B and is given to be $+1,-1,-1$ respectively. Moreover, the scalars giving Majorana and Dirac mass of RHNs carry same charge under $U(1)_X$ and $U(1)_{X'}$ symmetries  in both models. Therefore, the description of light neutrino mass matrix in both models is found to be same. Without loss of generality 
	we consider the model-A for the detailed study of generating sub-eV masses of light neutrinos. 
	
	The Lagrangian responsible for light neutrino masses in model-A is given by: 
	\begin{align}
		\mathcal{L} = & f_{ee} \overline{(N_e)^c} N_e \Phi + M_{e \mu} \overline{(N_e)^c} N_\mu + M_{e \tau} \overline{(N_e)^c} N_\tau + f_{\mu \mu} \Phi^\dagger \overline{(N_\mu)^c} N_\mu
		+ f_{\mu \tau} \Phi^\dagger \overline{(N_\mu)^c} N_\tau + f_{\tau \tau} \Phi^\dagger \overline{(N_\tau)^c} N_\tau \nonumber\\ 
		& + Y_{ee} \overline{L_e} \tilde{H_1} N_e + Y_{\mu \mu} \overline{L_\mu} \tilde{H_1} N_\mu + Y_{\mu \tau} \overline{L_\mu} \tilde{H_1} 
		N_\tau + Y_{\tau \mu} \overline{L_\tau} \tilde{H_1} N_\mu + Y_{\tau \tau} \overline{L_\tau} \tilde{H_1} N_\tau + h.c.
		\label{massnu}
	\end{align}
	
	After the $U(1)_X$ symmetry breaking, the right handed neutrino mass matrix $M_R$ and Dirac neutrino mass matrix $M_D$ have the form:
	
	\begin{equation}
		\label{mrmd}
		M_R=\left(
		\begin{array}{ccc}
			f_{ee} \langle \Phi \rangle & M_{e\mu} & M_{e\tau}\\
			M_{e\mu} & f_{\mu\mu} \langle \Phi \rangle & f_{\mu\tau} \langle \Phi \rangle\\
			M_{e\tau} & f_{\mu\tau} \langle \Phi \rangle & f_{\tau\tau} \langle \Phi \rangle
		\end{array}
		\right) ~~~~~~{\rm and}~~~~~~	M_D=\left(
		\begin{array}{ccc}
			Y_{ee} \langle H_1 \rangle & 0 & 0\\
			0 & Y_{\mu\mu} \langle H_1 \rangle & Y_{\mu\tau} \langle H_1 \rangle\\
			0 & Y_{\tau\mu} \langle H_1 \rangle & Y_{\tau\tau} \langle H_1 \rangle
		\end{array}
		\right)
	\end{equation}

	Using type I seesaw mechanism, the
	sub-eV light neutrino mass matrix can be found to be:
	\begin{equation}
		M_{\nu}=-M^T_{D} M^{-1}_R M_D\,,
	\end{equation}
	which has the texture:
	\begin{equation}
		M_\nu=\left(
		\begin{array}{ccc}
			A & B & C\\
			B & D & E\\
			C & E & F
		\end{array}
		\right)\,.
	\end{equation}
	The details of different entries of $M_\nu$ is given in Appendix~\ref{appenA}. We note that in our case both normal and inverted 
	mass hierarchies among light neutrinos are possible. This is in contrast to theories considering global 
	$L_e-L_\mu -L_\tau$ symmetry where the prediction for neutrino mass hierarchy is strictly inverted hierarchy.

	\section{DARK MATTER}\label{darkmatter}
	Due to the $Z_2$ symmetry in model-A, the $4^{th}$ generation neutrino becomes a potential DM candidate~\cite{Lee:2011jk,Borah:2011ve,Zhou:2011fr, Arina:2012aj}. The fourth generation neutrino combines with a gauge singlet right handed neutrino to form a heavy Dirac neutrino ~\cite{Hill:1989vn,Aparici:2012vx} and thus evades the constraint from the invisible $Z$ decay width from LEP measurements which strongly constrains the number of active light neutrinos to only 3. But such a dark matter candidate is ruled out by the relic density and direct detection constraints as demonstrated below. 
	
\begin{figure}[h!]
		\begin{center}
			\includegraphics[scale=0.5]{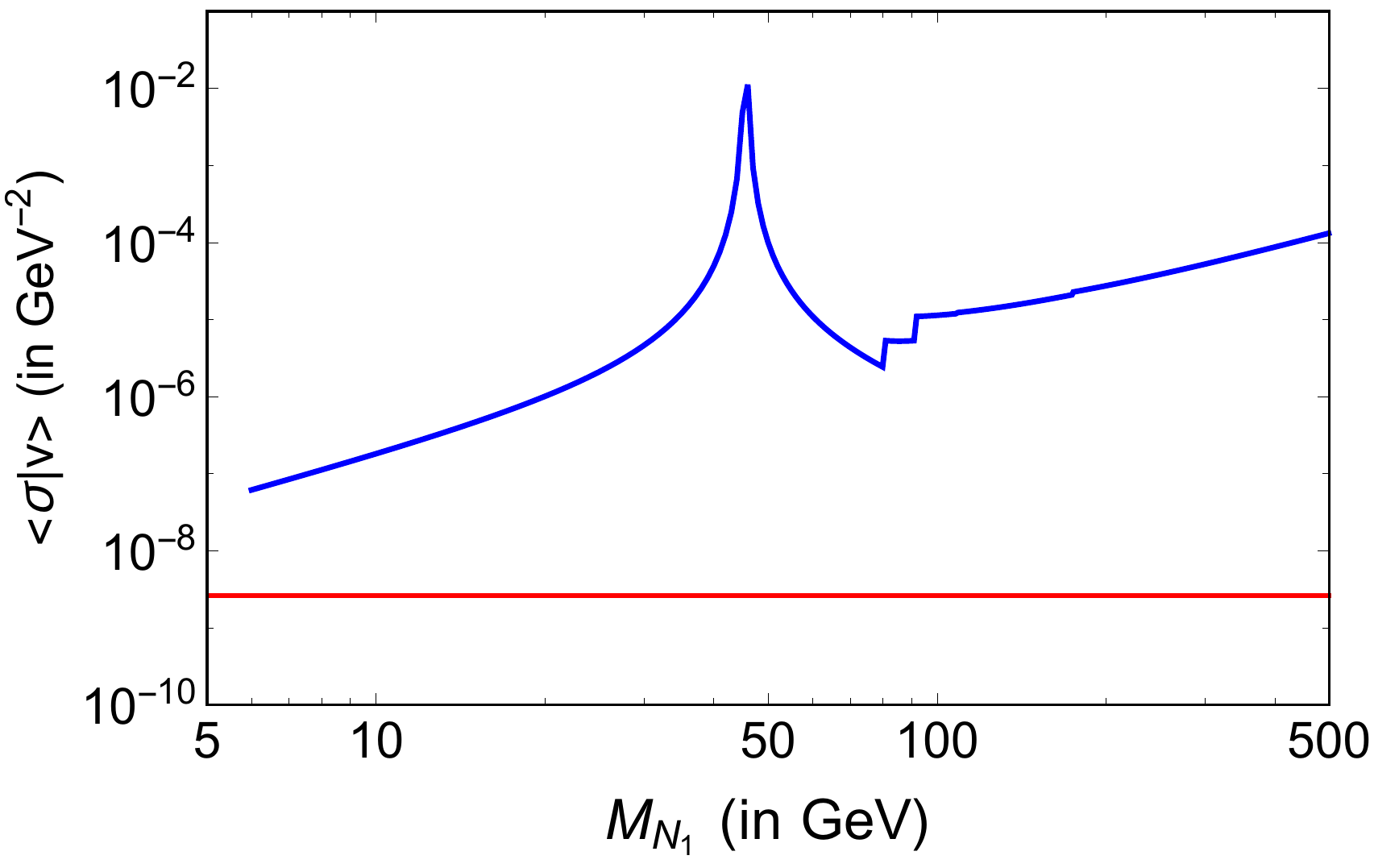}
			\caption{Thermally averaged annihilation cross-section times the relative velocity of $N_1$ plotted as a function of $M_{N_1}$. The red line depicts the $\langle \sigma v \rangle$ that can give rise to correct relic density of DM.}
			\label{taac}
		\end{center}
	\end{figure}

	The reason why such a DM candidate is ruled out is because of its large annihilation cross-section to SM particles. Therefore, it becomes under abundant and can at best give rise to 1$\%$ of the observed relic density. The thermally averaged cross-section of $N_1 N_1 \to SM ~SM$ is shown in Fig,~\ref{taac}, which clarifies this point. Moreover, because of its large coupling to the $Z$-boson, it can scatter off the nuclei inside terrestrial detectors via a tree level $Z$-boson exchange in direct detection experiments but the resulting cross-section is large enough that such DM particles would have already been detected if it were there as the present direct detection experiments like XENON1T has already probed the cross-section down to $10^{-47} {cm}^2$ ~\cite{Aprile:2018dbl}. Thus fourth generation neutrino can not be the single dominant component to account for the whole of DM, unless it is asymmetric inelastic \cite{Arina:2012aj}. 
	
	Therefore in the following we consider an alternative DM candidate. Due to our assumption that the tree-level mixing of the $U(1)_X$ breaking scalar $\Phi$ with $H_1$ and $H_2$ in model-A is zero, the real part of the $U(1)_X$ breaking scalar $\Phi$ (denoted as $\phi$) cannot decay at tree level vis Higgs mixing. Other tree level and a loop level decays of $\phi$ are discussed below and we show that, only if the DM mass is less than an MeV, the other decay modes are consistent with a lifetime long enough to be a dark matter of the Universe~\cite{Mohapatra:2020bze}.  Thus, the dark matter in our model is predicted to be a sub-MeV  dark matter. We discuss the various decay modes below  to show how this happens.
	
	Similarly in model-B, the real part of $U(1)_{X'}$ breaking scalar $\Phi_1$ (denoted as $\phi_1$) has similar features as that of $\Phi$ in model-A. Since we assume that the tree level mixing of $\Phi_1$, $\Phi_2$ with the SM Higgs H is zero, $\phi_1$ can also be a potential DM candidate in model-B. Thus the DM description in both models are same and need not be described individually. In the following, 
	we choose $\phi$ to be the candidate of dark matter in model-A for presentation purpose.   
	
	\vspace{1cm}
	\noindent \underline{\bf Lifetime Constraint}
	
\begin{figure}[h]
		\includegraphics[scale=0.8]{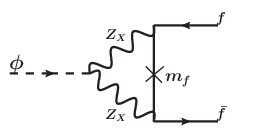}
		\caption{Feynman diagram for $\phi \to f \bar{f}$ via $Z_X$ loop.}
		\label{fig:phizploopdec}
	\end{figure}

	Now let us consider $m_\phi$ in the range $1$ MeV-$10$ GeV and analyze the viability of DM in this mass range. The channels through which $\phi$ can decay to SM fermions are:
	(I) $\phi\to N N \to l f \bar{f} l f \bar{f}$, (II) $\phi\to Z'Z'\to f \bar{f}f \bar{f}$
	and (III) $\phi \to f\bar{f}$ through mixing with SM Higgs boson at one-loop or via $Z'$ loop as shown in Fig.~\ref{fig:phizploopdec}. In the following, we analyse these three decay modes and find out the parameter space that is consistent with the lifetime constraint for the DM $\tau_\phi>10^{25}$ sec.
	
	The corresponding decay widths of these processes are given by:
	\begin{eqnarray}
		\Gamma(\phi\to NN \to \ell f \bar{f} \ell f \bar{f})&\simeq& \frac{1}{(4 \pi)^8} f^2 (	Y^2 \, y^2)^2 \frac{m^{13}_\phi}{M^4_N \, M_h^8}\label{decnn}\\
		\Gamma(\phi \to {Z_{_X}Z_{_X}}\to \ell  \bar{\ell} \ell  \bar{\ell})&\simeq& 
		\frac{g_{_X}^6}{ 256 \pi^5}  \frac{m_\phi^7}{M^6_{Z_{X}}}\label{deczxzx}\\
		\Gamma({\phi \to f \bar{f}}) &\simeq& \frac{1}{4 \pi} \theta^2 \left(\frac{m_f}{v} \right)^2 m_\phi \label{decloop}
	\end{eqnarray}
	where the mixing parameter $\theta$ is loop induced as the tree-level $H_{1,2}$-$\Phi$ coupling is set to zero to prevent  
	$h$ and $\phi$ from mixing at the tree level to ensure the lifetime of $\phi$ above the bound. This $H_{1,2}$-$\Phi$ coupling is absent in a supersymmetric embedding of this theory. and we assume that this can be done.
	Alternatively, if this mixing is set to zero at the tree level, it can arise from a RHN fermion box diagram at one-loop and  can be estimated to be: 
	\begin{eqnarray}
		\theta\sim \frac{f^2 Y^2}{16\pi^2}\frac{v v_{\phi}}{M^2_h} 
		\sim \frac{1}{16\pi^2}  \frac{m_\nu M_N^3}{v M_h^2} \frac{2 g_{_X}}{M_{Z_{X}}}
	\end{eqnarray}
	As has been shown in Ref~\cite{Mohapatra:2020bze}, for a lower than hundred GeV $M_N$, we find this one loop lower limit on $\theta$ to be consistent with lifetime constraints  for a $0.1$ GeV DM mass, and a low $g_X\sim 10^{-7}$ or so.

	Similarly, the decay width of $\phi$ to $f\bar{f}$ via the $Z'$ loop which is shown in Fig.~\ref{fig:phizploopdec}, is given by:
	\begin{equation}
		\Gamma(\phi\to f\bar{f}) \simeq \frac{g^6_x}{12 \pi (16 \pi^2)^2} \frac{m^2_f}{M^2_{Z_{X}}} m_\phi
		\label{eq:phideczploop}
	\end{equation}

	In Eq.~\ref{decnn}, Yukawa coupling $Y$ can be estimated from the neutrino mass requirement as $Y^2= \frac{m_\nu M_N}{v^2}$. Thus using typical neutrino mass scale $m_\nu \simeq 0.1$ eV and for a fixed benchmark value of $M_N$, the constraint on the mass of DM $m_\phi$ and the product of Yukawa couplings $f y^2$ can be found out from the DM lifetime bound. If $\phi$ mass lies in the MeV scale , then for a $\mathcal{O}(10^2)$ GeV scale $M_N$ the lifetime of $\phi$ translates  to $\tau_{\phi}\sim8.9\times10^{73}{~\rm s}/(fy^2)^2$ . Thus this constraint is quite consistent with the requirement for $\phi$ to be the DM.
	
	%
	The second decay mode shown in Eq.~\ref{deczxzx} is most crucial from the phenomenological point of view as it involves the gauge coupling and gauge boson mass and thus puts stringent constraint on these two parameters which are also crucially relevant for generating the DM relic as we will discuss in the next section. 
	
	
	

	The third decay channel mentioned in Eq.~\ref{decloop}, is possible through the $\Phi-H_{1,2}$ mixing. But to achieve the long lifetime of $\phi$ from the DM requirements, we set their tree-level coupling $\lambda_{1\Phi} (H_1^\dagger H_1) (\Phi^\dagger\Phi)$ and $\lambda_{2\Phi} (H_2^\dagger H_2) (\Phi^\dagger\Phi)$ to zero. Thus the loop induced coupling through the RHN loop comes into play and it can be small enough to make the lifetime of $\phi$, $\tau_\phi$ greater than $10^{25}$ s. Here, it is worth mentioning that if $m_\phi$ is greater than the SM Higgs mass $M_h$, then $\phi$ can decay to a pair of $h$ and the loop induced mixing is not small
	enough to be consistent with the lifetime constraint. Thus this restricts one to keep $m_\phi$ light. 
	
	
	As mentioned earlier, another crucial decay channel is $\phi \to f \bar{f}$ via the $Z_{X}$ loop as shown in Fig.~\ref{fig:phizploopdec} and the corresponding constraint for the lifetime of $\phi$ coming from this decay mode turns out to be the most stringent one as we will discuss in the next section. 
	
	Here, we note that there is also a two neutrino decay mode of $\phi$ {\it i.e.} $\phi \to N N \to \nu \nu$ with each $N$ mixing with the light neutrinos via Dirac mass. This decay width of $\phi$ decaying to two neutrinos is given by:
	\begin{equation}
		\Gamma(\phi \to N N \to \nu \nu) = \frac{1}{4\pi}\left(\frac{M_N}{v_\phi}\right)^2 \left(\frac{m_\nu}{M_N}\right)^2  m_\phi\,
	\end{equation}
	and we estimate that, for $m_\phi=1$ MeV,$M_N=100$ GeV and $M_N/v_\phi \leq 10^{-5}$, this decay width is less than $10^{-41}$ GeV or so giving a lifetime longer than the age of the Universe.

	%

	\vspace{1cm}
	
	\noindent\underline{\bf Relic Density of DM}
	
	As discussed in the previous section, to ensure the long lifetime of $\phi$ such that it qualifies to be the DM candidate, the values of the gauge coupling $g_{_X}$ and gauge boson mass $M_{Z_X}$ are restricted in a particular range for each value of $m_\phi$. As $\phi$ remains out of equilibrium with the SM bath for values of $g_X$ that satisfy the lifetime constraint, the typical thermal freeze-out mechanism for producing DM relic density becomes inapplicable, and one must instead investigate the freeze-in mechanism.
	
	For this, we can rely on the gauge boson annihilation ({\it i.e.}$Z_X Z_X \to \phi \phi$) to generate the relic density of $\phi$ via freeze-in. This in turn requires $Z_X$ to be in equilibrium with the SM bath. 
	As discussed in~\cite{Mohapatra:2019ysk}, the most efficient process for such gauge boson $Z_{X}$ to be in equilibrium with SM particles is via the process $f\bar{f}\to Z_{X}+\gamma$ and the condition on $g_{X}$ for this to happen  is  $g_{X} > 2.7\times 10^{-8}\left({M_{Z_{X}}}/{{\rm 1\,  GeV}}\right)^{1/2}$. 
	
	\begin{figure}[h]
		\includegraphics[scale=0.5]{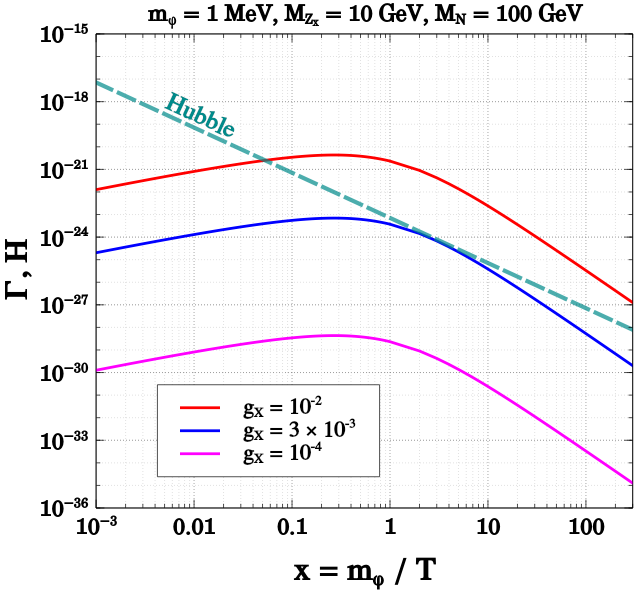}
		\caption{Comparison of the interaction rate $\Gamma = n \langle \sigma v \rangle$ with the Hubble expansion rate(dashed line) for the processes $Z_{_{X}} Z_{_{X}} \to \phi \phi$ (solid lines) for different values of the gauge couplings.}
		\label{intrateH}
	\end{figure}
	
	Another upper bound on $g_{X}$ comes from the requirement that the DM particle $\phi$ is out of equilibrium in the early universe.
	For this, we consider the two processes  $Z_{X} Z_{X} \leftrightarrow \phi \phi$ and $NN \leftrightarrow \phi \phi$ which might bring $\phi$ to equilibrium with the SM bath.
	As the out-of-equilibrium condition is given by $n \langle \sigma v \rangle < H$ where  the Hubble parameter $H =\sqrt{\frac{\pi^2}{90} g_*} T^2/M_{Pl.}$ with the reduced Planck mass $M_{Pl.}=2.43 \times 10^{18}$ GeV 
	and the effective total number of relativistic degrees of freedom $g_*=106.75$, in Fig.~\ref{intrateH}, we have shown the interaction rate $Z_{X} Z_{X} \leftrightarrow \phi \phi$ against the Hubble expansion rate for three different values of $g_X$ for $m_{\phi}=1$ MeV. Clearly, $g_X > 3 \times 10^{-3}$ brings the DM to equilibrium and hence is not viable. We note that for $M_{N}$ of $\mathcal{O}(10^2)$ GeV, the corresponding $\Gamma(NN \leftrightarrow \phi \phi)$ remains sub-dominant as compared to $\Gamma(Z_X Z_X \leftrightarrow \phi \phi)$. 
	
	

	\begin{figure}[h]
		\includegraphics[scale=0.5]{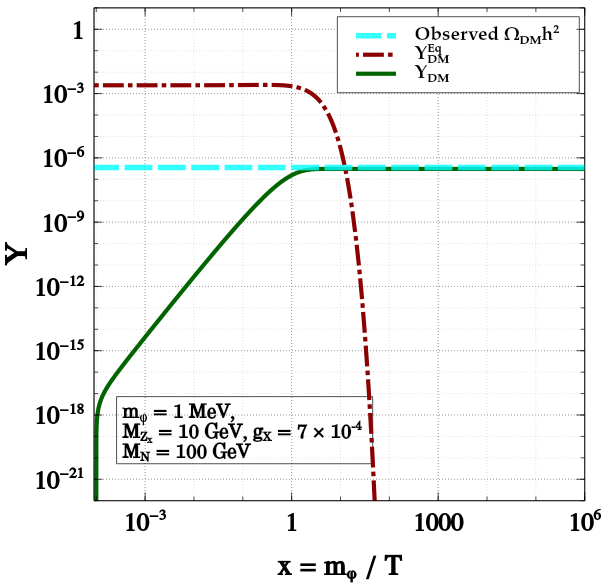}
		\caption{Evolution of the comoving number density of the DM via freeze-in for typical benchmark values of the parameters as mentioned in the inset of the figure. }
		\label{Ydmevo}
	\end{figure}
	
	\vspace{1cm}
	We solve the
	following Boltzmann equation to evaluate the relic density of $\phi$ produced via the freeze-in mechanism as discussed above:
	\begin{equation}
		\frac{dY_\phi}{dx} = \frac{s(m_\phi)}{x^2 H(m_\phi)} \langle \sigma v \rangle (Y^{\rm Eq}_{\phi})^2
	\end{equation}
	where $Y_\phi$ is the comoving number density of $\phi$, $x=m_{\phi}/T$, $Y^{\rm Eq}_\phi$ is the comoving number density of $\phi$ if it were in thermal equilibrium. $s(m_\phi)$ and $H(m_\phi)$ are the entropy density and Hubble parameter respectively which are given by: $s(m_{\phi}) = \frac{2\pi^2}{45} g_{*s} m^3_\phi$ and $H(m_\phi)=1.67 g^{1/2}_* \frac{m^2_\phi}{M_{Pl.}}$. In Fig.~\ref{Ydmevo}, we demonstrate the evolution of comoving number density of $\phi$ for a fixed set of benchmark values of the parameters that generates the correct relic density of $m_\phi=1$ GeV DM. 
	\begin{figure}[h]
		\includegraphics[scale=0.5]{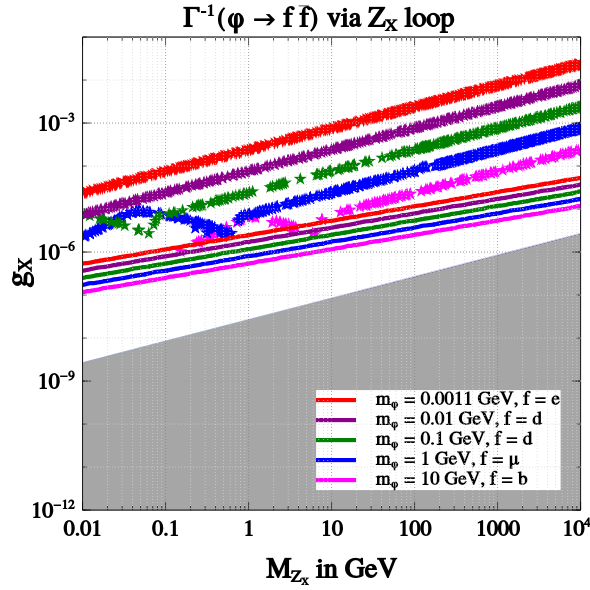}
		\caption{Relic density satisfying points in the $g_X - M_{Z_X}$ plane for different values of $m_{\phi}$. The coloured star-shaped points satisfy 
			the relic density constraint. The coloured straight lines represent the upper limit on $g_X$ from the long lifetime requirement of DM.}
		\label{psf}
	\end{figure}
	
	In Fig.~\ref{psf}, we have shown the points satisfying correct relic density for different values of $m_\phi$ which are depicted by the coloured diamonds. The grey line depicts the lower bound on $g_X$ from the requirement of keeping $Z_X$ in equilibrium with the thermal bath and the other solid lines are the most stringent life time constraints for different $m_\phi$ coming from the $\phi$ decay to fermions via the $Z_X$ loop as mentioned in Eq.~\ref{eq:phideczploop}. Here, we note that the other constraints mentioned in Eq.~\ref{decnn}, \ref{deczxzx} and \ref{decloop}, and are somewhat less stringent as compared to the constraint from Eq.~\ref{eq:phideczploop}.
	
	Here it is worth noticing that, this lifetime constraint is extremely severe and thus  completely rules out the parameter space satisfying correct relic density. Hence in our scenario GeV or sub-GeV scale DM is not viable as they can decay to SM charged fermions and the stringent constraint on DM lifetime $\tau_{DM}>10^{25}$ s coming from the Gamma ray observations is applicable~\cite{Baring:2015sza}.
	
	\begin{figure}[h]
		\includegraphics[scale=0.45]{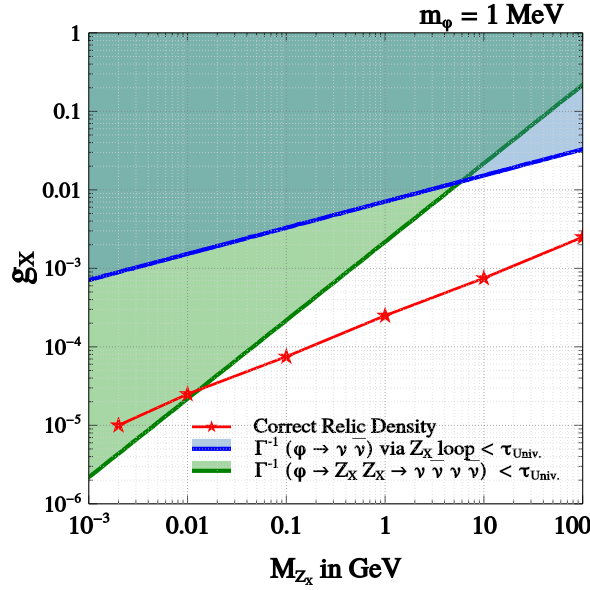}
		\hfil
		\includegraphics[scale=0.45]{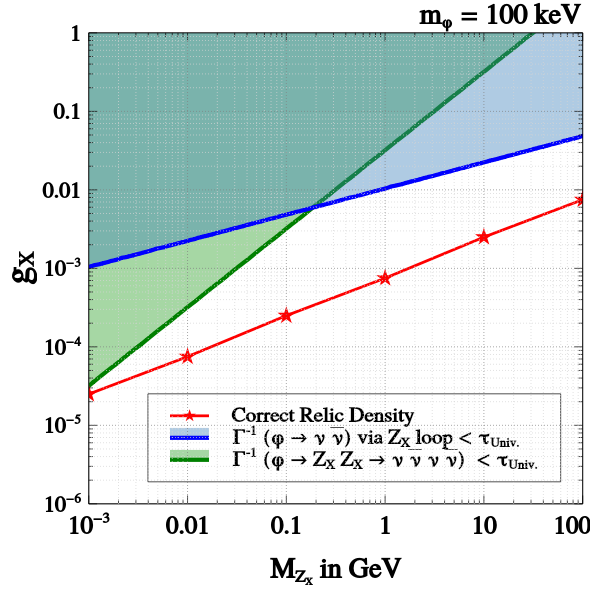}
		\caption{Parameter space in the plane of $g_X-M_{Z_X}$ plane consistent with relic and lifetime constraints of sub-MeV DM for $m_{\phi}=1$ MeV and $m_\phi=100$ keV. }
		\label{wdm_bp_summary}
	\end{figure}

	Therefore we resort to sub-MeV scale DM where the only channel possible is $\phi$ decaying to neutrinos. This has two advantages and interesting phenomenological consequences. As in such a scenario the decay width is suppressed by $m_\nu$ squared, hence one can obtain a sufficiently large lifetime for the DM $\phi$ and since there are no experimental constraints on such light DM decaying to neutrinos, satisfying the lifetime constraint $\tau_{\rm DM} > \tau_{\rm Univ}$ suffices the purpose. Here $\tau_{\rm Univ.}$ is the age of the Universe which is $\tau_{\rm Univ.}\sim 5 \times 10^{17}$ s.
	
	In Fig.~\ref{wdm_bp_summary}, we showcase the parameter space consistent with both relic density and lifetime requirement for DM masses $m_{\phi}=100$ keV and $m_{\phi}=1$ MeV. Here the relic satisfying points are depicted in red colour and the most stringent lifetime constraints has been shown by the shaded regions. Clearly, we can find a viable parameter space consistent with both correct relic density and DM lifetime criteria.


	\vspace{1cm}
	\noindent\underline{\bf Direct Search Constraint}
	
\begin{figure}[h]
		\includegraphics[scale=0.15]{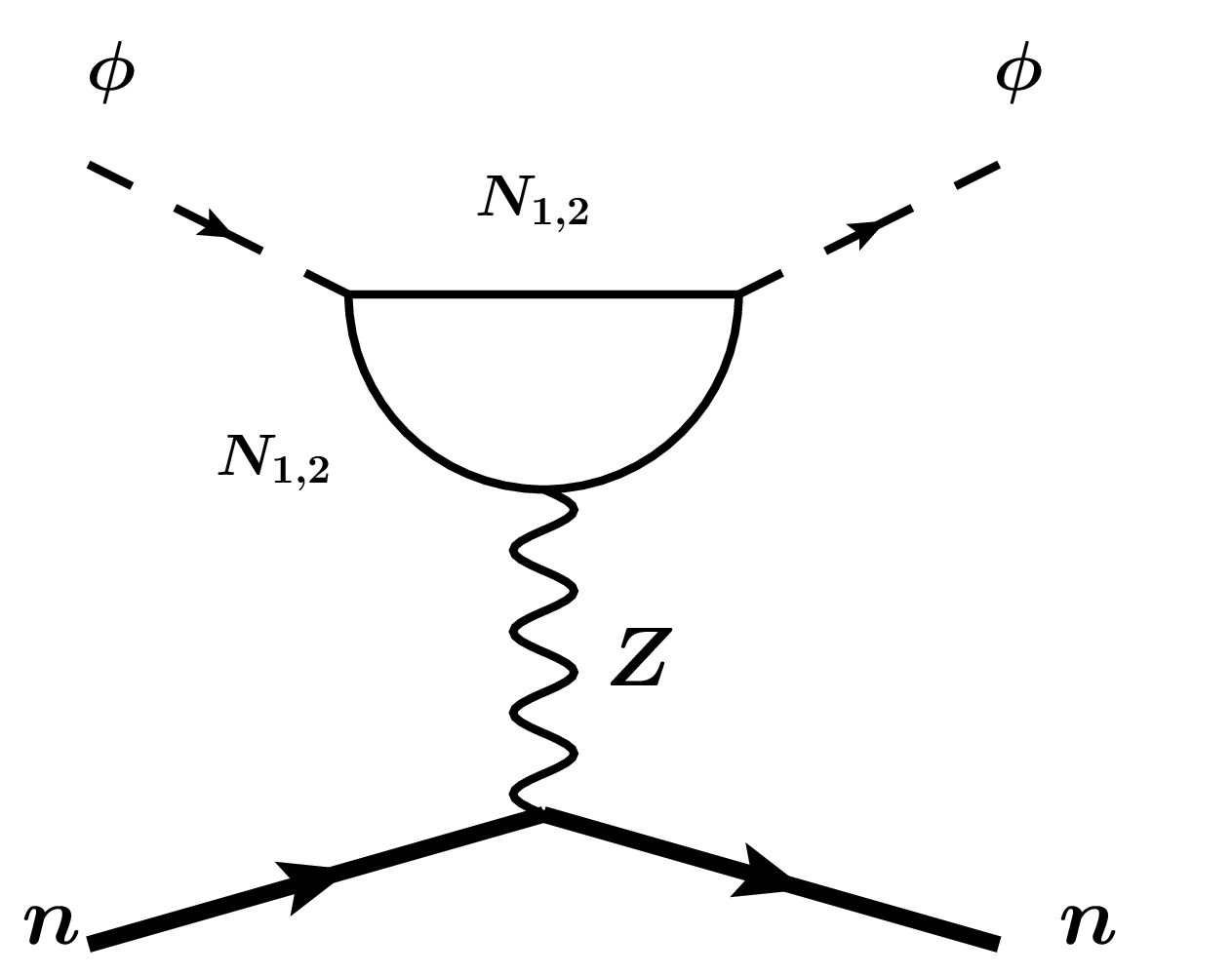}
		\caption{DM-nucleon scattering arising at one-loop level with the 4th generation neutral leptons in the loop.}
		\label{loopdd}
	\end{figure}

	In this scenario, as the DM $\phi$ is a singlet scalar and since its tree level coupling with SM Higgs has been set to zero to rescue it from the lifetime constraints, DM-nucleon scattering is not possible at tree level to be probed by the terrestrial DM direct search experiments. However, at one loop level, DM can scatter off the nuclei as shown in Fig.~\ref{loopdd} with the fourth generation neutral leptons in the loop.  
	
	However, this DM-nucleon elastic scattering is spin-dependent for which the direct search constraints are comparatively less stringent as compared the spin-independent DM-nucleon elastic scattering. Moreover because of additional loop suppression with the heavy RHNs in the loop, this cross-section is very small and hence easily evades the direct detection constraints.
	Here it is worth mentioning that, the diagram shown in Fig.~\ref{loopdd}, is for Model-A. For Model-B, the respective diagram for DM-nucleon scattering is mediated by $Z_{X'}$.

		\section{Anomalous Magnetic Moment of Leptons}
		\label{g-2anomaly}

		
		The recent measurement of muon anomalous magnetic moment by the E989
		experiment at Fermi lab 
		$a^{\rm FNAL}_\mu = 116 592 040(54) \times 10^{-11}$ 
		when combined with the previous Brookhaven determination of
		$a^{\rm BNL}_\mu = 116 592 089(63) \times 10^{-11}$
		leads to a 4.2 $\sigma$ observed excess of
		$\Delta a_\mu = 251(59) \times 10^{-11}$ over the SM prediction ($
		a^{\rm SM}_\mu = 116 591 810(43) \times 10^{-11}
		$)\cite{Abi:2021gix}.
		
		Similarly anomalous magnetic moment for electron measurement using Rubidium atomic interferometry gives us $+$ve value of $\Delta a_e$ with $1.6\sigma$ discrepancy with SM predicted value \cite{Morel:2020dww}
		$(\Delta a_e)_{\rm Rb} = ( 48 \pm 30) \times 10^{-14}.
		\label{del_ae}$

		In the models under consideration, this additional contribution to muon and electron magnetic moment can come from the one-loop diagram mediated by $Z_{X}$ gauge boson. This one-loop contribution is given by \cite{Brodsky:1967sr, Baek:2008nz}
		\begin{equation}
			\Delta a_{l} = \frac{\alpha_X}{2\pi} \int^1_0 dx \frac{2m^2_{l} x^2 (1-x)}{x^2 m^2_{l}+(1-x)M^2_{Z_{X}}} \approx \frac{\alpha_X}{2\pi} \frac{2m^2_{l}}{3M^2_{Z_{X}}}
		\end{equation}
		where $\alpha_X=g^2_{X}/(4\pi)$ and $m_l$ is the mass of the corresponding lepton.

		\begin{figure}[h]
			\includegraphics[scale=0.5]{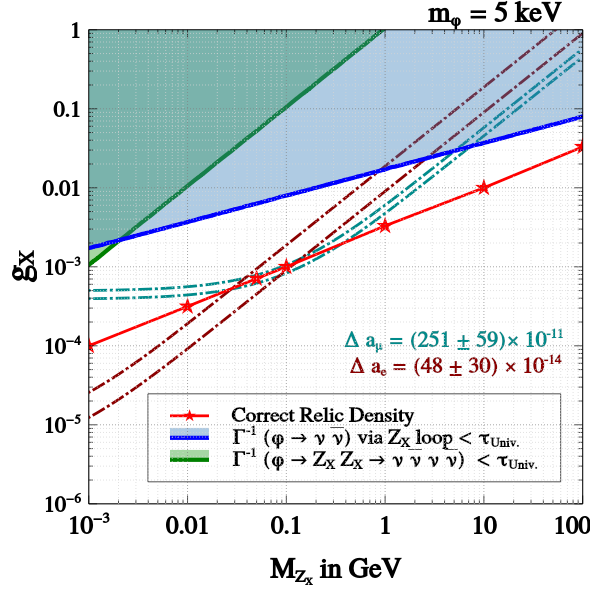}
			\caption{Parameter space in the plane of $g_X-M_{Z_{X}}$ plane for $m_\phi=5$keV}
			\label{wdm_summary}
		\end{figure}
		
		In Fig.~\ref{wdm_summary}, we showcase the parameter space for a benchmark value of the DM mass $m_\phi=5$ keV, that is consistent with both lifetime and relic abundance constraints as well as can explain the muon and electron $(g-2)$ simultaneously. Thus, from this analysis we also infer that, to explain both electron and muon $(g-2)$ the allowed values of $g_X-M_{Z_X}$ can successfully generate the correct relic density for light DM {\it i.e $m_\phi=5$} keV.

		\section{COLLIDER SIGNATURES}\label{collider_sign}
		\begin{figure}[h]
			\includegraphics[scale=0.15]{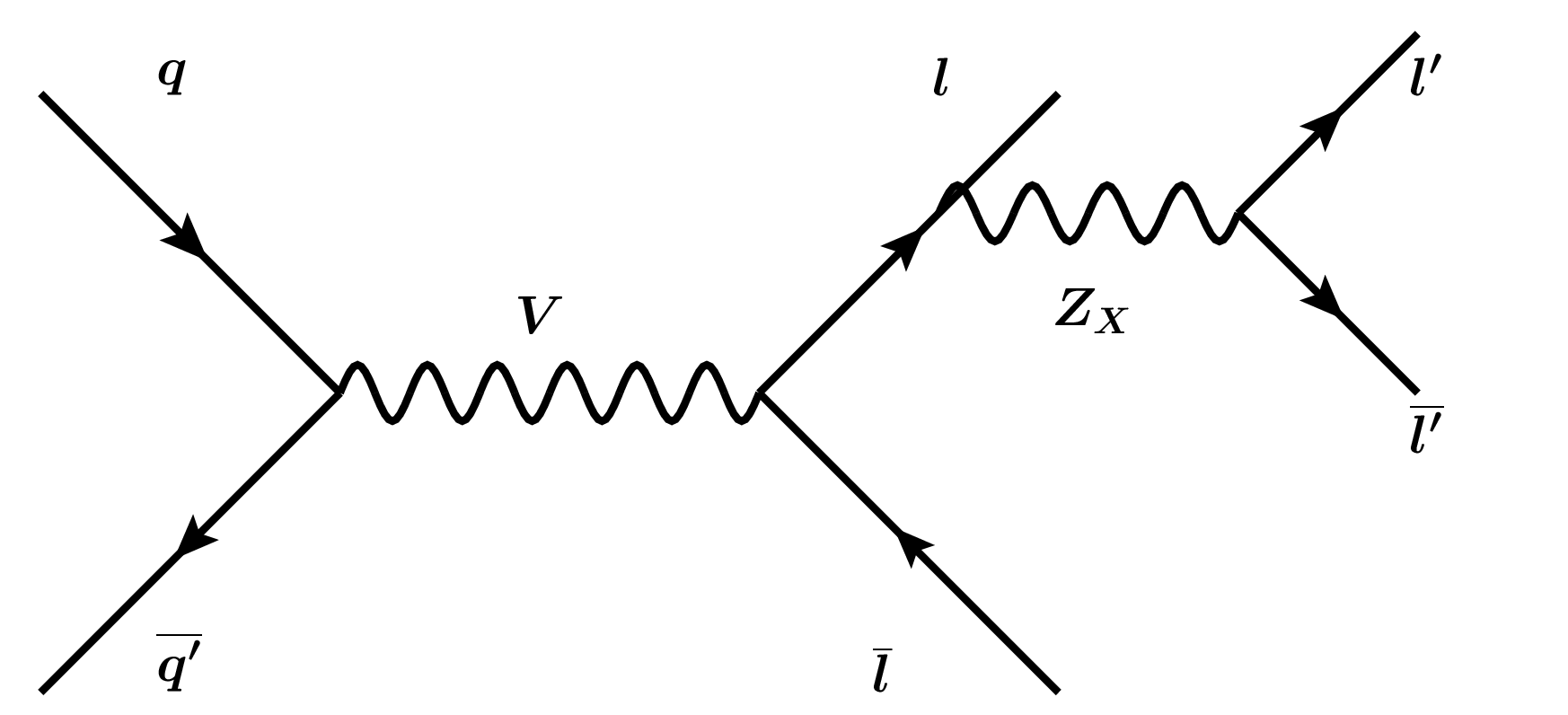}
			\caption{Production diagram of $Z_X$ at collider experiments.}
			\label{collider}
		\end{figure}
		
		This kind of gauged Leptonic symmetries can have interesting signatures at the collider. Especially for model-A, the production of such leptophilic gauge boson $Z_{X}$ in the presence of the 4th generation of leptons can give promising prospects of discovery in the present and future generation colliders~\cite{delAguila:2014soa,CMS:2018yxg}. The production of $Z_{X}$ at the LHC is illustrated in Fig.~\ref{collider}, where $Z_X$ is emitted from one of the final leptons in Drell-Yan production. This process can give rise to the four-lepton signals and if the exchanged electroweak gauge boson is $W^{\pm}$, then one can see the three charged leptons plus missing energy signal. Here it is worth mentioning that, in model-B, the gauge boson $Z_{X'}$ behaves identical to the $\rm B-L$ gauge boson.
		
		Apart from the gauge bosons, another interesting possibility arises for model-B, through which it can be probed at colliders via the displaced vertex signature of the scalar $\Phi_2$ which can decay to two quarks through its coupling mentioned in Eq.~\ref{quarkmassB} or to two leptons through its mixing with SM Higgs.

		\section{CONCLUSION}\label{conclusion}
		In this paper, we discussed two models, namely model-A and model-B, based on the gauged $L_e-L_\mu-L_\tau$ extension of 
		the SM. In model-A, the gauge anomaly cancellation requires that there be four generation of quarks and 
		leptons in which case, the gauge group becomes $L_e-L_\mu-L_\tau+L_4$. In order to generate light neutrino masses through type-I seesaw, we further introduced four generations of 
		right handed neutrinos $N_e, N_\mu, N_\tau$ and $N_4$ having $L_e-L_\mu-L_\tau+L_4$ charges 1,-1,-1,1 respectively. An SM singlet 
		scalar $\Phi$ having $L_e-L_\mu-L_\tau+L_4$ charge -2 was introduced to break the gauged $L_e-L_\mu-L_\tau+L_4$ symmetry at a TeV scale. 
		We also introduced two Higgs doublets $H_1$ and $H_2$ to give masses to SM particles. By imposing a $Z_2$ symmetry 
		we restricted $H_2$ to give masses to down type particles while $H_1$ to that of up type particles. Similarly by 
		imposing another $Z_2$ symmetry we segregated the 4th generation of particles so that the existing constraints can be evaded. The first $Z_2$ symmetry is anyway required to make the heavy fourth generation compatible with the 
		oblique parameter constraints from S and T.
		
		Similarly in model-B, we considered a gauged $(L_e-L_\mu-L_\tau)-(B_1-B_2-B_3)$ extension of the SM with $B_a$ 
		representing the $a^{\rm th}$ generation of quarks. The model was anomalous within the particle content of the SM. 
		However, we made it anomaly free by introducing three generation of right handed neutrinos $N_e, N_\mu, N_\tau$ 
		with $(L_e-L_\mu-L_\tau)-(B_1-B_2-B_3)$ charges 1,-1,-1 respectively. This model works for three fermion generations. Unlike the model-A, in this case, we introduced 
		two SM singlet scalars $\Phi_1$ and $\Phi_2$ with $(L_e-L_\mu-L_\tau)-(B_1-B_2-B_3)$ charges 2 and 2/3 respectively 
		to break the symmetry. The vev of $\Phi_1$ generates $3\times 3$ light neutrino mass matrix with non-zero entries through type-I seesaw 
		mechanism, while vev of $\Phi_2$ give masses to quarks. In this set up, we find that the role of $\Phi_1$ mimics the role of 
		$\Phi$ in model-A.   
		
		We showed that there are two important consequences of having gauged $L_e-L_\mu-L_\tau$ symmetry: (1) The vev of $\Phi$ generates 
		$3\times 3$-Majorana mass matrix of RHNs with non-zero entries, which results in a light neutrino mass matrix having 
		both possibilities of normal hierarchy and inverted hierarchy. We note that models with global 
		$L_e-L_\mu-L_\tau$ symmetry predicts only an inverted hierarchy pattern of light neutrino masses. (2) By assuming 
		the tree-level mixing between $\Phi-H_1$ and $\Phi-H_2$ (in model-A) to be zero, we showed as in Ref.~\cite{Mohapatra:2020bze} that the real part of $\Phi$ 
		(called as $\phi$ here) can be a viable dark matter candidate. Since $\phi$ is unstable, we derived the relevant 
		constraints for it to qualify as dark matter. After imposing all the relevant constraints, we find that this scenario dictates for a sub-MeV scale DM to give rise to correct relic density that is consistent with the DM lifetime constraints as shown in Fig.~\ref{wdm_bp_summary}. We also showcase the parameter space that can explain both electron and muon anomalous magnetic moment simultaneously while successfully explaining the DM of the Universe consistent with the lifetime and relic density constraints.

		\section*{Acknowledgements}\label{acknowlwdgw}
		NS acknowledges the support from Department of
		Atomic Energy (DAE)- Board of Research in Nuclear
		Sciences (BRNS), Government of India (Ref. Number:
		58/14/15/2021- BRNS/37220). RNM would like to thank Kaustubh Agashe for useful discussions.
%
		
		\appendix
		\section{Neutrino Mass Matrix}
		\label{appenA}
		The elements of the light neutrino mass matrix are: 
		\begin{eqnarray}
			A&=& -\frac{p^2 \left(e^2-d f\right)}{a \left(e^2-d f\right)+b^2 f-2 b c e+c^2 d}\\
			B&=&\frac{p (b e s-b f q-c d s+c e q)}{a \left(e^2-d f\right)+b^2 f-2 b c e+c^2 d}\\
			C&=&\frac{p (b e t-b f r-c d t+c e r)}{a \left(e^2-d f\right)+b^2 f-2 b c e+c^2 d}\\
			D&=&\frac{a \left(d s^2-2 e q s+f q^2\right)-b^2 s^2+2 b c q s-c^2 q^2}{a \left(e^2-d f\right)+b^2 f-2 b c e+c^2 d}\\
			E&=&\frac{a (d s t-e (q t+r s)+f q r)+b^2 (-s) t+b c (q t+r s)-c^2 q r}{a \left(e^2-d f\right)+b^2 f-2 b c e+c^2 d}\\
			F&=&\frac{a \left(d t^2-2 e r t+f r^2\right)-b^2 t^2+2 b c r t-c^2 r^2}{a \left(e^2-d f\right)+b^2 f-2 b c e+c^2 d}
		\end{eqnarray} 
		
		where the $M_R$ and $M_D$ in Eq.~\ref{mrmd}, are written as: 
		\begin{equation}
			M_R=\left(
			\begin{array}{ccc}
				a & b & c\\
				b & d & e\\
				c & e & f
			\end{array}
			\right) ~~~~~~{\rm and}~~~~~~	M_D=\left(
			\begin{array}{ccc}
				p & 0 & 0\\
				0 & q & r\\
				0 & s & t
			\end{array}
			\right)
		\end{equation}
		Clearly, this mass matrix $M_\nu$ can lead to either normal or inverted mass ordering for neutrinos.
		
\providecommand{\href}[2]{#2}\begingroup\raggedright\endgroup


	\end{document}